\documentclass{iaus}
\usepackage{graphicx}

\title[Grain growth in protoplanetary disks] 
{3D SPH simulations of grain growth in protoplanetary disks}

\author[Guillaume Laibe \& Jean-Fran\c ois Gonzalez]   
{Guillaume Laibe$^1$,
Jean-Fran\c cois Gonzalez$^1$,
Laure Fouchet$^2$ and
Sarah T. Maddison$^3$}

\affiliation{$^1$Universit\'e de Lyon, Lyon, F-69003, France; Universit\'e Lyon 1, Villeurbanne, F-69622, France; CNRS, UMR 5574, Centre de Recherche Astrophysique de Lyon,\\ \'Ecole Normale Sup\'erieure de Lyon, 46 all\'ee d'Italie,
F-69364 Lyon cedex 07, France\\
email: {\tt Jean-Francois.Gonzalez@ens-lyon.fr, Guillaume.Laibe@ens-lyon.fr} \\[\affilskip] $^2$Department of Physics, ETH Zurich, CH-8093 Zurich, Switzerland\\
email: {\tt fouchet@phys.ethz.ch}
\\[\affilskip]
$^3$Centre for Astrophysics and Supercomputing, Swinburne University of Technology,\\ PO Box 218, Hawthorn, VIC 3122, Australia\\
email: {\tt smaddison@swin.edu.au}}

\pubyear{2008}
\volume{249}  
\jname{Exoplanets: Detection, Formation and Dynamics}
\editors{A.C. Editor, B.D. Editor \& C.E. Editor, eds.} 
\begin{document}

\maketitle

\begin{abstract}
We present the first results of the treatment of grain growth in our 3D, two-fluid (gas+dust) SPH code describing protoplanetary disks. We implement a scheme able to reproduce the variation of grain sizes caused by a variety of physical processes and test it with the analytical expression of grain growth given by Stepinski \& Valageas (1997) in simulations of a typical T Tauri disk around a one solar mass star. The results are in agreement with a turbulent growing process and validate the method. We are now able to simulate the grain growth process in a protoplanetary disk given by a more realistic physical description, currently under development. We discuss the implications of the combined effect of grain growth and dust vertical settling and radial migration on subsequent planetesimal formation.
\end{abstract}

\firstsection 
\section{Introduction}
\label{Sectintro}

Collisions and aggregation govern the 
first steps of planet formation from micron-sized particles to decimetric pre-planetesimals.
Observations of protoplanetary disks support this
mechanism by showing evidence of dust grain growth (e.g. Apai et al.\ 2004).
We describe here the implementation of grain growth
in our simulations of protoplanetary disks.
Depending on their relative
velocities and material properties (Dominik \& Tielens 1997), collisions
between solid particles can make them stick and grow, or conversely break
them into smaller pieces.

\section{Grain growth in protoplanetary disks}
\label{SectSV97}

A first description of the grain growth process is given by Stepinski \&
Valageas (1997). They model a turbulent, vertically isothermal
protoplanetary disk, in which gas and dust are represented by two separate
phases interacting via aerodynamic drag in the Epstein regime. Their solid
particles are supposed to stick perfectly during collisions, and can therefore
only grow. The variation of their size $s$ is given by the following
analytical expression of the growth rate:
\begin{equation}
\frac{\mathrm{d}s}{\mathrm{d}t}=\sqrt{2^{^3\!/\!_2}\,\mathrm{Ro}\,\alpha}\,
\frac{\rho_\mathrm{s}}{\rho_\mathrm{d}}\,C_\mathrm{s}
\frac{\sqrt{\mathrm{Sc}-1}}{\mathrm{Sc}},
\label{EqEvol}
\end{equation}
where Ro is the Rossby number for turbulent motions, $\alpha$ the Shakura \&
Sunyaev (1973) viscosity parameter, $\rho_\mathrm{s}$ the density of matter
concentrated into solid particles, $\rho_\mathrm{d}$ the intrinsic density of
the grains, $C_\mathrm{s}$ the local gas sound speed, and Sc the Schmidt
number of the flow which estimates the effect of gas turbulence on the grains.
Sc is defined by
\begin{equation}
\mathrm{Sc}=(1+\Omega_\mathrm{k}\,t_\mathrm{s})
\sqrt{1+\frac{\bar{\mathbf{v}}^2}{V_\mathrm{t}^{2}}},
\label{EqSchmidt}
\end{equation}
where $\Omega_\mathrm{k}$ is the local keplerian velocity, $t_\mathrm{s}$
the dust stopping time, $\bar{\mathbf{v}}$ the mean relative velocity between
gas and dust, and $V_\mathrm{t}$ a turbulent velocity. The growth rate
depends on $s$ through the stopping time
\begin{equation}
t_\mathrm{s}=\frac{\rho_\mathrm{d}\,s}{\rho_\mathrm{g}\,C_\mathrm{s}},
\label{EqStopTime}
\end{equation}
where $\rho_\mathrm{g}$ is the gas density.

The motion of a dust grain and the
grain growth process, which both depend on the grain
size $s$, are coupled phenomena. The understanding of the global evolution of
dust in disks therefore requires a numerical treatment.

\section{Grain growth with an SPH code}

Our 3D, bi-fluid, Smooth Particle Hydrodynamics (SPH) code has been developed
 to model vertically isothermal, non self-gravitating
protoplanetary disks. Gas and dust are treated as two
separate phases and are coupled by aerodynamic drag. The code and its results
on dust migration and settling are presented in Barri\`ere-Fouchet et al.\
(2005). It has also been applied to grain stratification in GG~Tau's
circumbinary ring (Pinte et al.\ 2007) and to gaps opened by planets in the
dust phase of protoplanetary disks (Maddison et al.\ 2007; Fouchet et al.\
2007; see also talk by Gonzalez, this volume).

The assumptions we have made are very similar to those of Stepinski \&
Valageas (1997) mentioned in Sect.~\ref{SectSV97}, their prescription for
grain growth is therefore easy to implement in our code. 
We allow the grain size $s$ assigned to each SPH particle,
assuming it to represent the typical size of dust grains at its position, to
vary with time following Eq.~(\ref{EqEvol}). We take the initial grain size
distribution to be uniform.

\section{Results}

\begin{figure}[here]
\begin{center}

\begin{minipage}[c]{.46 \linewidth}
\begin{center}
\includegraphics[height=5.6cm,angle=-90]{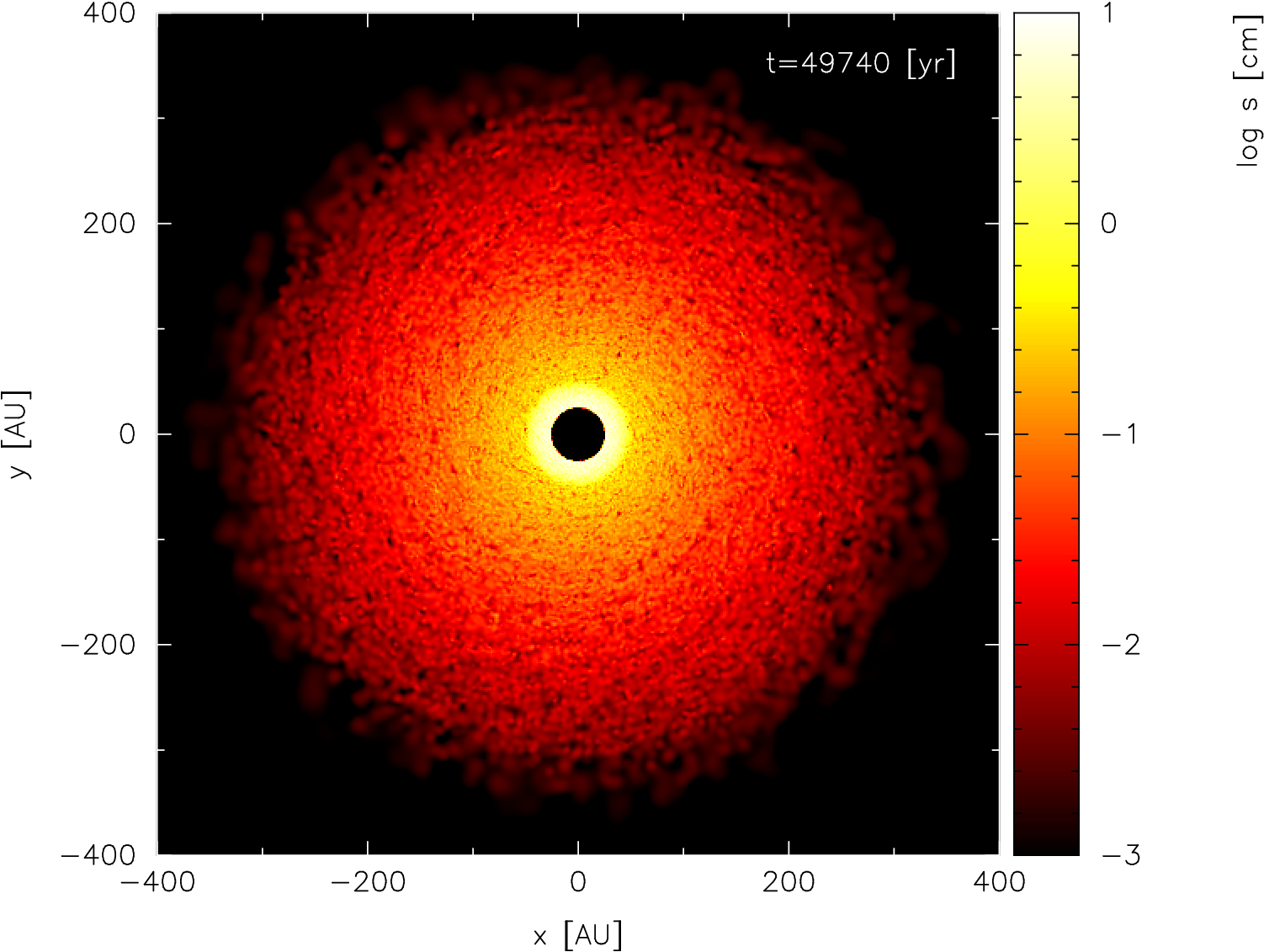} 
\caption{Grain size distribution in an equatorial plane cut of the disk.}
\label{Figdiskxy}
\end{center}
\end{minipage} \hfill
\begin{minipage}[c]{.46 \linewidth}
\begin{center}
\includegraphics[height=5.6cm,angle=-90]{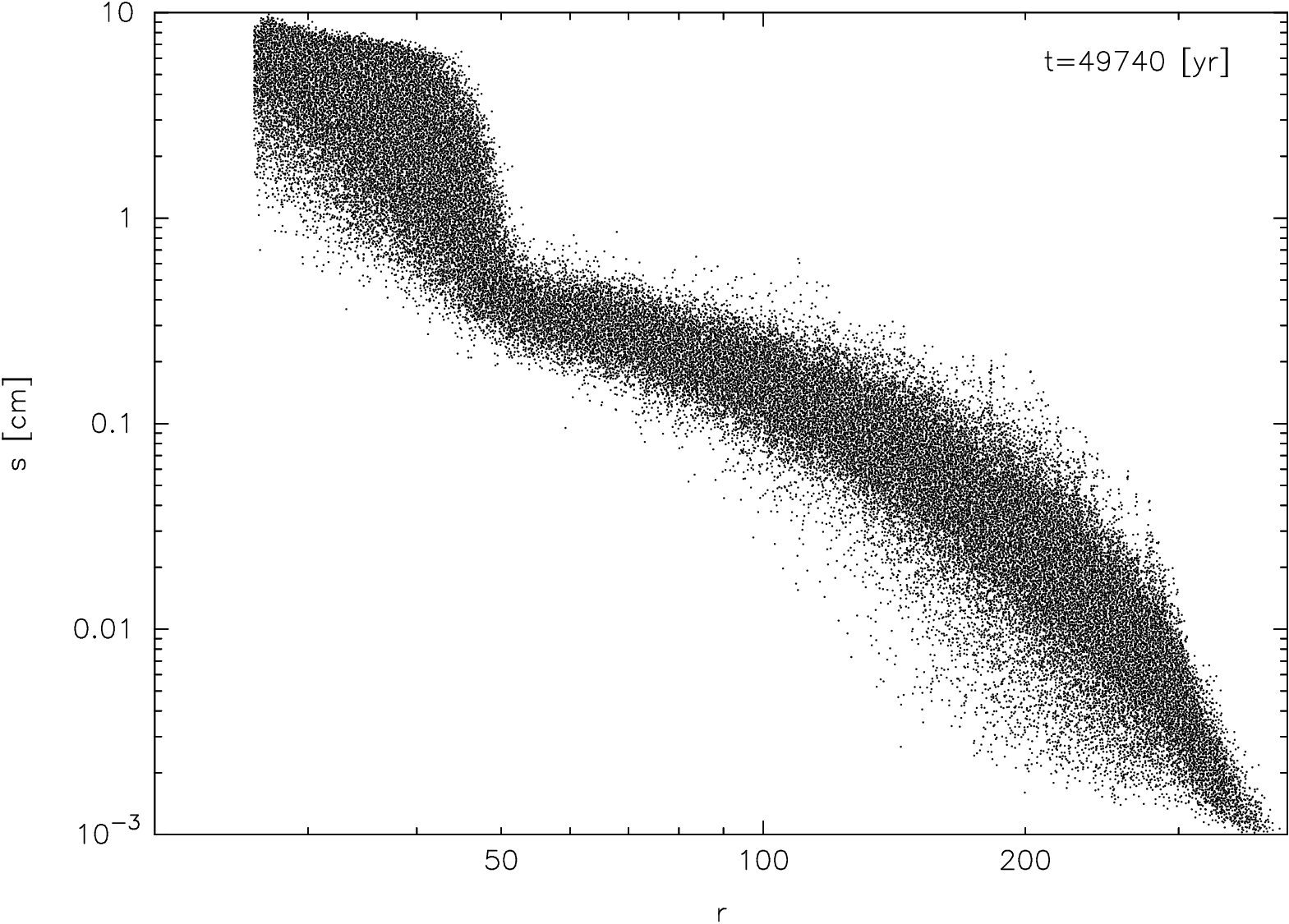} 
\caption{Grain size distribution as a function of radial distance to the star.}
\label{Figdisksr}
\end{center}
\end{minipage}
\begin{minipage}[c]{.46 \linewidth}
\medskip
\begin{center}
\includegraphics[height=5.6cm,angle=-90]{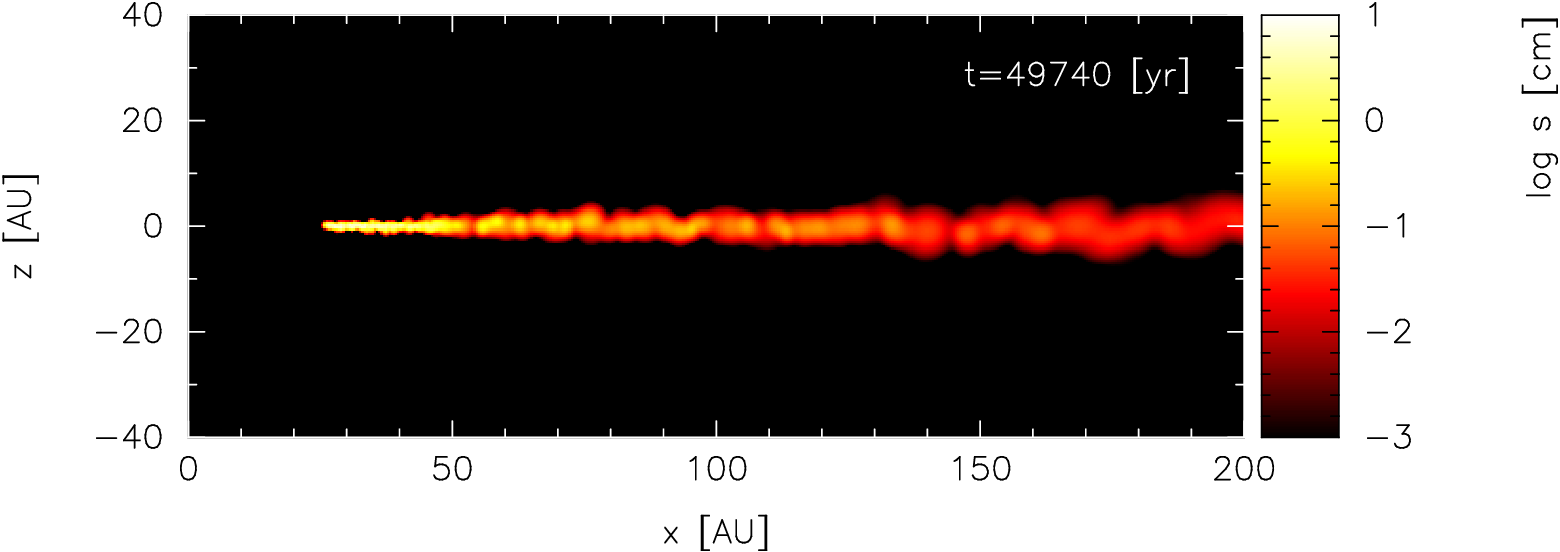} 
\end{center}
\end{minipage} \hfill
\begin{minipage}[c]{.46 \linewidth}
\medskip
\begin{center} 
\caption{Grain size distribution in a meridian plane cut of the disk. Figures are made with SPLASH (Price 2007).}
\label{Figdiskxz}
\end{center}
\end{minipage} \hfill

\end{center}
\end{figure}

We model grain growth in a typical T~Tauri disk of mass $M_\mathrm{disk}=0.01\
M_\odot$, with a total dust mass $M_\mathrm{dust}=0.01\ M_\mathrm{disk}$,
around a central star of mass $M_\star=1\
M_\odot$. We ran simulations with 200,000 SPH particles, evolved over 50,000
years, for a series of initial dust grain sizes ranging from $s_0=1\ \mu$m to
1~mm.
Fig.~\ref{Figdiskxy} and Fig.~\ref{Figdiskxz} show the resulting grain sizes
in face-on and edge-on views of the disk, for $s_0=1\ \mu$m.

We find that grain growth occurs very quickly, especially in the inner disk
where the density is the highest,
as one would expect from Eq. (\ref{EqEvol}). In this region, the dust grains
reach centimetric size (see Fig. \ref{Figdisksr}) in only a few timesteps. In
the outer parts of the disk, where the density is far lower, the grains grow much more slowly and their size stays
below the millimeter.

An inwards migrating grain coming into a denser region grows almost
instantaneously to a size characteristic of the grains at its new position,
which in turn evolves slowly over time. Hence, the global shape of the plot shown
in Fig.~\ref{Figdisksr} stays the same as time goes on,
but the size distribution slowly progresses to larger sizes.
The initial size $s_0$ has very little influence on the final size
distribution, and in particular on the maximum size reached.

Our results are consistent with those obtained by Dullemond \& Dominik (2005).
With a model solving the coagulation equation in presence of turbulence, they
found a very fast grain growth in T~Tauri protoplanetary disks, depleting very
small sizes and producing centimetric grains.

\section{Conclusion}

We have implemented in our 3D SPH code a mechanism able to treat grain
growth and validated it through the use of the simple model of Stepinski \&
Valageas (1997).

In accordance with physical intuition, dust grains grow much more quickly in the
denser, central regions of the disk, where centrimetric sized are reached.
The growth time is smaller than the migration time and a quasi-stationnary
distribution of grain size appears in the disk. The small grains are depleted
too rapidly to be consistent with observations of protoplanetary disks,
showing the need to take into account other processes such as microscopic interactions
between the grains, kinetic energy dissipation, porosity, and re-fragmentation. This will be addressed in future work.

\end{document}